# How the Saha Ionization Equation Was Discovered

Arnab Rai Choudhuri

Department of Physics, Indian Institute of Science, Bangalore – 560012

## Introduction

Most youngsters aspiring for a career in physics research would be learning the basic research tools under the guidance of a supervisor at the age of 26. It was at this tender age of 26 that Meghnad Saha, who was working at Calcutta University far away from the world's major centres of physics research and who never had a formal training from any research supervisor, formulated the celebrated Saha ionization equation and revolutionized astrophysics by applying it to solve some long-standing astrophysical problems.

The Saha ionization equation is a standard topic in statistical mechanics and is covered in many well-known textbooks of thermodynamics and statistical mechanics [1–3]. Professional physicists are expected to be familiar with it and to know how it can be derived from the fundamental principles of statistical mechanics. But most professional physicists probably would not know the exact nature of Saha's contributions in the field. Was he the first person who derived and arrived at this equation? It may come as a surprise to many to know that Saha did not derive the equation named after him! He was not even the first person to write down this equation! The equation now called the Saha ionization equation appeared in at least two papers (by J. Eggert [4] and by F.A. Lindemann [5]) published before the first paper by Saha on this subject. The story of how the theory of thermal ionization came into being is full of many dramatic twists and turns. The aim of the present article is to elucidate the nature of Saha's contributions to this subject and to discuss the circumstances which led to his extraordinary works. While preparing this article, I greatly benefitted from a perusal of the outstanding popular account of Saha's science given by Venkataraman [6]. A few other authors also have given accounts of the Saha ionization equation [7–10].

## Some background information about Saha

A full, authoritative biography of Saha still remains to be written. However, a broad outline of his life can be found in the biographical accounts of Saha that we have [6, 11–13]. Although very little archival materials about Indian scientists of that era have been preserved [14], luckily Saha's personal papers – especially his correspondence with other scientists – have survived, thanks to the efforts of his children [15].  Some of Saha's personal letters to fellow scientists (especially to H.N. Russell and H.H. Plaskett) give us a rare glimpse of the inner workings of the brilliant mind that came up with the theory of thermal ionization.

Son of a small grocery shop owner, Meghnad was born on 6 October 1893 in the nondescript village of Seoratali not too far from Dacca.  Although there was no academic atmosphere in the immediate family surroundings, he grew up to be an exceptional student.  After completing high school and intermediate from Dacca, Meghnad joined Presidency College in Calcutta in 1911.  There he obtained BSc degree in Mathermatics and MSc degree in Mixed Mathematics (this curious name was changed to Applied Mathematics in 1936) in the years 1913 and 1915 respectively. Satyendra Nath Bose, who was later to formulate the Bose-Einstein statistics, was Saha's classmate in Presidency College and secured the first position both in the BSc and MSc examinations.  Saha came out second.

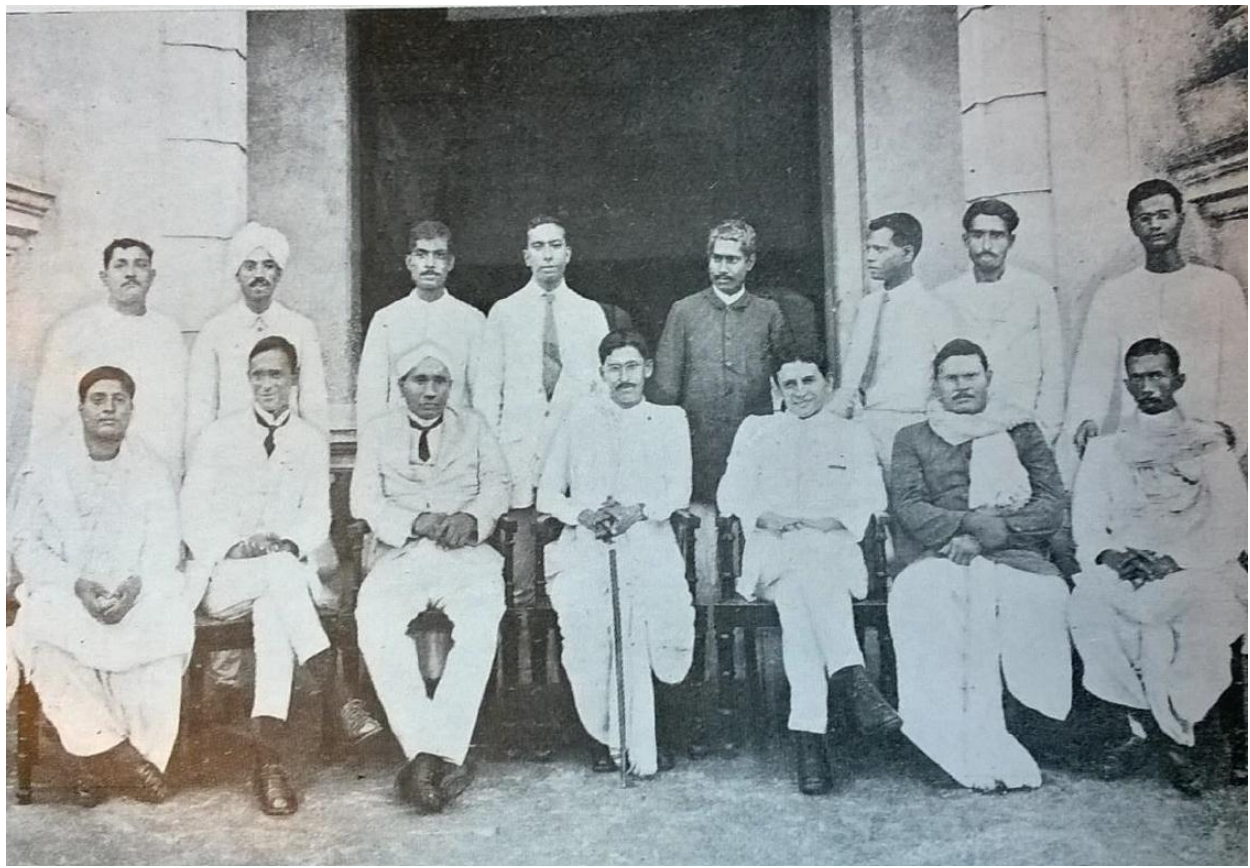

Fig. 1.  A group photograph of the physics faculty of Calcutta University taken on the eve of Saha's departure to Europe in 1920.  Several well-known physicists can be seen.  Seated: S.N. Bose (1st from left), C.V. Raman (3rd from left), M.N. Saha (4th from left), D.M. Bose (5th from left); Standing: S.K. Mitra (4th from left).

When Saha and Bose graduated with MSc degree in 1915, the job market in Calcutta for science graduates was not particularly bright. But an unusual opportunity suddenly opened up. Till that time, Calcutta University was merely a body for regulating and conducting examinations without any core faculty in any academic discipline. Sir Asutosh Mookerjee, whose term as Vice-Chancellor of Calcutta University had ended in 1914 (he would later be Vice-Chancellor for another term), was still the de facto Vice-Chancellor and was in the process of establishing various academic departments with core faculty. Mookerjee was a capable mathematician himself, who could not continue the mathematical researches which he started in his youth due to lack of opportunities. He had a particularly soft corner for physics and knew of the pathbreaking developments taking place in physics at that time. A physics department was planned in the newly established University College of Science. Sir Asutosh Mookerjee wanted it to be a place where teaching and research on latest developments of physics would be carried out. He was on the lookout for the best available faculty. S.N. Bose had given an account of the meeting which Sir Asutosh had with him, Saha and their batchmate Sailen Ghosh, who was the class topper in MSc Physics ([16], my translation from Bengali) : "The three of us – I, Meghnad and Sailen, overwhelmed with awe and reverence – climbed the steep stairs to reach the office of Sir Asutosh. He had heard that the youngsters were willing to teach new topics at the university. He asked us, 'Boys, what will you be able to teach?' 'We shall try to teach whatever you assign to us.' Asutosh smiled. . . Meghnad was assigned to teach quantum theory, whereas the responsibility of teaching Einstein's Relativity Theory was given to me. We agreed on the condition that we would be given one year's time for preparation."

The Physics Department of Calcutta University College of Science started in 1916 with a small core faculty of barely half a dozen persons. The most prestigious chair – the Palit Professorship – was offered to C.V. Raman, who was for a few years posted in Calcutta as an officer in the Finance Department and was hand-picked by Sir Asutosh, who knew about the research Raman was doing in the evenings and in the weekends. As is well known, three extraordinary physics discoveries came from people who constituted the small core faculty of this new Physics Department – the Saha inonization equation in 1920, the Bose statistics in 1924 (although Bose had shifted to Dacca University by that time, he continued to have close links with Calcutta University) and the Raman effect in 1928. Our concern in the present article is with the first of these extraordinary discoveries.

## Saha's intellectual preparation

The normal procedure of getting initiated into academic research is to enroll under a supervisor, who would usually ask the student to concentrate on a focused subject. Usually the student would be expected to master the subject of research in which the supervisor has been active and then continue investigations in that subject in a focused manner. This certainly did not happen in the case of Saha and Bose, who had taken up teaching positions at Calcutta University after their MSc. At that time, Raman was in the process of building up his research group by attracting many students. Both Saha and Bose received hints that it would be in their best interests to enroll for research under Raman. However, Raman's initial reputation was based primarily on his works on acoustics and optics, which did not interest Saha or Bose. They were eager to do something in what they perceived as the frontier of physics at that time. So they had to find their own path. There are indications that Saha initially might have tried to work with Raman, but things did not work out. S.N. Bose, who must have observed this interaction closely, commented many years later ([16], my translation from Bengali): "Meghnad did not have good experimental skills, which displeased the Palit Professor." Saha's own account of this is, however, different (Saha to H.H. Plaskett, 21 December 1946 [17]): "I found that it is believed outside that during some early part of my career, I had been Raman's pupil or scholar. This is completely incorrect; I never owed anything to him in life, except persistent ill-will and attempt to harm me whenever possible."

The titles of Saha's first few papers make it abundantly clear that initially he was trying to find a path for himself by working on problems in very diverse areas of physics: "On Maxwell's stresses", "On the limit of interference in the Fabry-Perot interferometer", "On a new theorem in elasticity", "On the pressure of light", "On the dynamics of the electron", "On the influence of the finite volume of molecules on the equation of state" (with S.N. Bose, this being Bose's first paper), "On the fundamental law of electrical action" [18]. As the titles suggest, many of these papers were immediate extensions of what could be found in standard textbooks of advanced physics. While most of these papers were theoretical papers, the work on the pressure of light was experimental. At the end of this paper, we find the following acknowledgement: "we beg to record our best thanks to Prof. C.V. Raman." Although several of these early papers appeared in leading journals like *Philosophical Magazine* and *Physical Review*, it took Saha some time (about 3 years!!!) before he would stumble upon a niche area in which he could quickly become the world leader.

Saha has given an account of his intellectual preparation in a letter to H.H. Plaskett written many years later [17]: "I began from 1916 to read rather desultorily any book on Physics, Mathematics and Astronomy and Astrophysics, which came in my way or could be found in the library of my old college (Presidency College, Calcutta), or in the library of the Calcutta University. I might add that I had begun to learn German privately as early as 1911, while I was a student in the Inter science class, and by 1916 I had enough proficiency to study scientific papers without the use of a dictionary. In course of these studies I came across Miss Agnes Clerke's two books on astrophysics – one on the Sun, the other on Stars – and these excited my interest in Astrophysics, and made me familiar with some of its problems. A year later in 1917, the Calcutta University opened M.Sc. classes in Mathematics and Physics and I was asked to teach an odd assortment of subjects: Thermodynamics, Spectroscopy, Figure of the Earth, and was given charge of the Heat Laboratory. I was asked to teach thermodynamics because no one else of my colleagues would agree to take up that unpleasant subject, as they styled it. I had read no book on this subject previously."

Perhaps this habit of reading "rather desultorily" prepared Saha for the extraordinary synthetic work that he was to undertake soon. Saha's great work required the knowledge of several completely disconnected scientific fields – chemical equilibrium theory, atomic physics experiments, stellar spectroscopy. A young person trained under a regular research supervisor from an early age in a conventional manner would normally be expected to 'specialize' in a focused field and would not have time to acquire the knowledge of several such disconnected scientific fields. It is quite probable that at that time nobody in the whole world except Saha knew sufficiently about these disconnected fields to realize that results from these fields could be combined together to solve some of the frontier problems of science of that age. It is often pointed out that the lack of a formal training occasionally helps creativity. The case of the young Saha is a striking example of this.

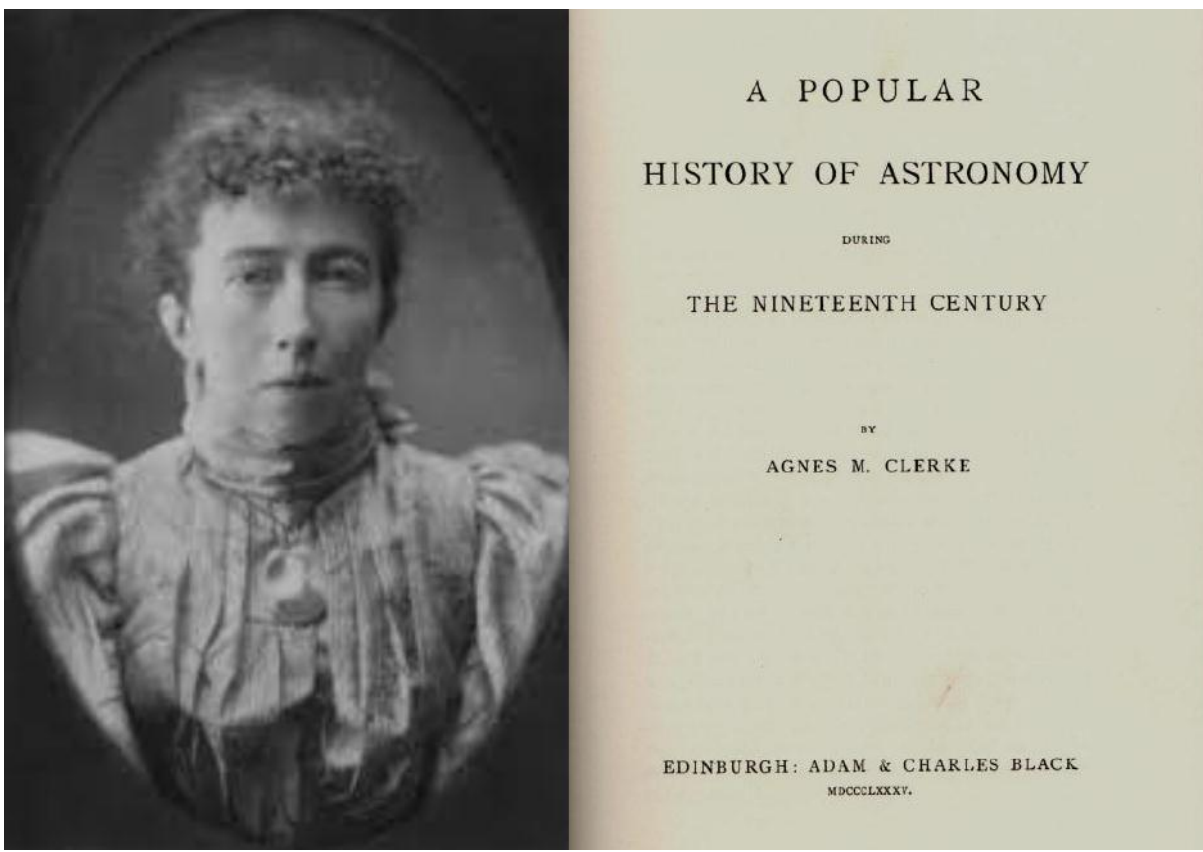

Fig. 2. Miss Agnes Mary Clerke (1842–1907) and the title page of her most famous book published in 1885. The young Saha learnt astrophysics from her books.

It may not be too out of place to make a few comments about Saha's training in astrophysics – the field which he was destined to revolutionize. As Saha himself mentioned, he picked up astrophysics from the books of Agnes Clerke, the exceptionally gifted popularizer of astronomy in Victorian England. We would, however, be missing the point if we think of Clerke's books as popular science books in the modern sense. She lived an era when astrophysics had not yet become very professionalized and it was still possible for a truly exceptional writer to discuss the most recent advances of astrophysics in books meant for the general reader. There had not been any popular books of astronomy of that class after Agnes Clerke! A perusal of her books leaves very little doubt that she could have become one of the leading astronomers of her age, if she was allowed to pursue a professional career in astronomy. But doing research in astronomy in that era usually meant staying up through the long night hours in an observatory far away from all residential areas to cut down stray light. The Victorian society could hardly be expected to allow a woman to follow this vocation. So Clerke channelized her unfulfilled desire of becoming an astronomer into truly extraordinary books meant for the general reader. It is good luck that her books fell into the hands of the young Meghnad, who had no opportunity for a professional training in astrophysics. Agnes Clerke was his true teacher in astrophysics. Perhaps there is symbolic significance in the fact that a young man at the periphery of the mighty British empire – who was disadvantaged because of his race – would learn astrophysics from a woman who – although close to the power centre of the empire – was disadvantaged because of her gender.

## Saha's scientific predecessors

After Rutherford established the nuclear model of the atom in 1911, it did not require much ingenuity to guess the possibility of a particular state of matter in which some electrons had come out of atoms. The astrophysical significance of this was apparent to Sir Arthur Eddington, who was working on the problem of stellar structure at that time. The average density of the Sun is a few times the density of water. One may naively expect the materials in the solar interior to have sufficiently high densities to keep them in the solid or liquid state. Now, we know that the properties of solids and liquids are much more complicated than the properties of gases. If materials at the centre of the Sun existed in the solid state (which was once believed to be the case), then it would have been very hard to make a theoretical model of the Sun. It was Eddington who realized that, if matter inside stars obeyed the ideal gas law, then only it would be possible to make realistic theoretical models of stars. Certainly, if the material existed in the plasma state, then this would be the case. Eddington [19] wrote in 1917 of "the hypothesis that the atoms are highly ionized, so that most or all of the electrons outside the nucleus have been broken off and move as independent particles. This suggestion that at these high temperatures we are concerned with particles smaller than the atom was made to me independently by Newall, Jeans, and Lindemann. By an argument which now appears insufficient, I had supposed that the atomic disintegration, though undoubtedly occurring, could not have proceeded very far; but Jeans has convinced me that a rather extreme state of disintegration is possible, and indeed seems more plausible." The challenge was to demonstrate through a quantitative rigorous calculation that atoms inside stars were really ionized. How could this be done?

The first person to realize that the theory of chemical equilibrium showed the path ahead was John Eggert [4], a physical chemist who was to become a pioneer of colour photography in later life. Suppose a calcium atom breaks into a calcium ion and an electron by absorbing energy $U$:

$$Ca + U \leftrightarrow Ca_+ + e. \quad (1)$$

This equation is clearly similar to the chemical equation representing chemical substances A, B, C, … reacting together to produce chemical substances M, N, O, …, i.e

$$A + B + C + … + U \leftrightarrow M + N + O + …, \quad (2)$$

where $U$ is the heat of reaction. This reaction would be reversible if all the substances involved are either (i) dissolved in a dilute solution, or (ii) exist in gaseous state inside a container. Then no product can leave the system and we expect to arrive at a chemical equilibrium in which the rate of the reaction in one direction balances the rate in the opposite direction, so that the concentrations of various chemical substances no longer change with time. The conditions for such chemical equilibrium were investigated by Gibbs. If temperature and pressure are held fixed, we expect the chemical potential to be extremum when equilibrium is reached. Suppose $\mu_A$ is the chemical potential of the substance A, and so on. Then the variation of the total chemical potential due to the reaction (2) will be zero if

$$\mu_A + \mu_B + \mu_C + … = \mu_M + \mu_N + \mu_O + … \quad (3)$$

If we can use the basic principles of statistical mechanics to obtain the µ-s in terms of such quantities as temperature, pressure and the concentrations of the various chemical substances, then we get an equation relating the concentrations of the reacting chemical substances under the equilibrium condition.

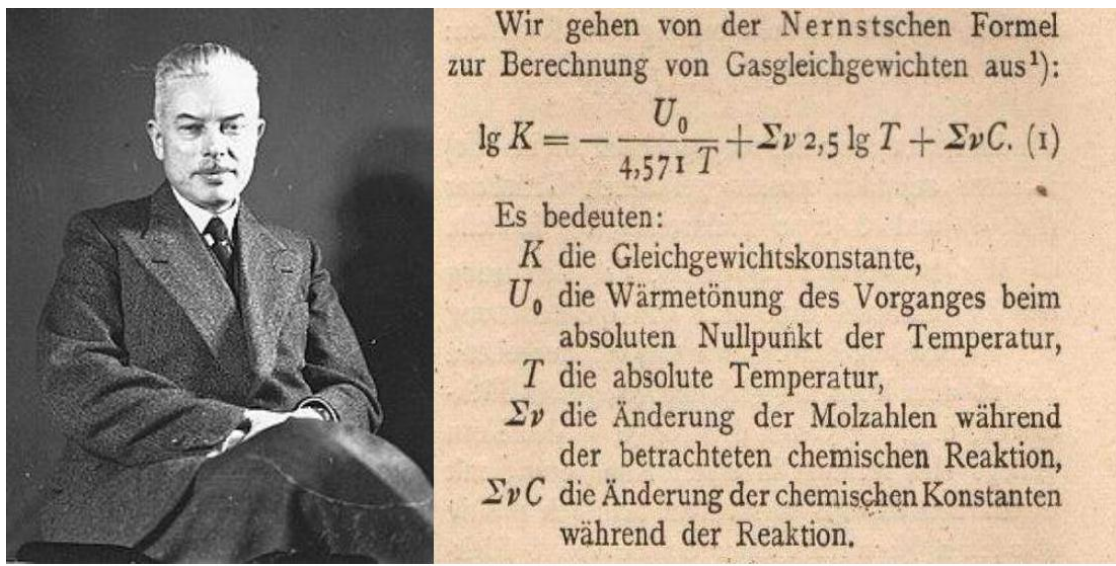

Fig. 3. John Eggert (1891 – 1973) and the part of his paper [4] where the ionization equation was written down.

One important comment needs to be made here about the application of this theory to gaseous reactions. In order to develop a quantitative theory, it is necessary to do a summation over all possible states of the particles, atoms or molecules. This is done by dividing the phase space into cells representing individual states and then integrating over the phase space. In classical physics, the division of the phase space into cells representing individual states is an arbitrary process. We need to apply quantum principles to conclude that a volume $h^3$ of the six-dimensional position-momentum phase space corresponds to a single state. In modern textbooks on statistical mechanics, usually the Heisenberg uncertainly principle is invoked to justify this. However, more than a decade before the uncertainly principle was formulated, O. Sackur and H.M. Tetrode independently discovered this important result in 1912. The so-called Sackur-Tetrode formula is essential for doing any quantitative calculation in this subject.

Eggert [4] was the first person to point out in his paper "Über den Dissoziationszustand der Fixsterngase" ("On the State of Dissociation in the Inside of Fixed Stars" – according to Saha's translation of the title) that the theory of chemical equilibrium could be applied to the problem of ionization. Eggert's mentor Walther Nernst [20] had written down a formula for chemical equilibrium in 1918 in his classic treatise on the heat theorem, which now proved particularly convenient. Fig. 3 shows the part of Eggert's paper in which he writes down what he calls "Nernstschen Formel" (Nernst's formula) in a form suitable for application to study ionization. Anybody familiar with the Saha ionization equation will immediately recognize that this is nothing other than that equation written in the general logarithmic form. Now, to do any quantitative calculation with this equation, one needs the value of the heat of reaction denoted by the symbol $U_0$ by Eggert. He correctly surmised that $U_0$ for the ionization of hydrogen would be given by the energy required to remove the electron from the Bohr orbit inside the hydrogen atom. But Eggert had no idea how to obtain $U_0$ for any other chemical element. So he could do some quantitative estimates only for hydrogen and argued that hydrogen would be ionized inside stars, thereby lending support to Eddington's theory.

Had it been realized at that time that hydrogen is the most abundant chemical element inside stars, then perhaps Eggert's study would have appeared more compelling. But this was first established in Cecilia Payne's famous PhD thesis of 1925, when the theory of thermal ionization had been fully developed and Payne was able to apply this to the relevant data of stellar spectra. At the time of Eggert's work, it was widely believed by astronomers that the composition of the Sun is not too different from that of the Earth. So the burning question was to do ionization calculations for other elements, to check whether the Sun is really ionized to a high degree. This was accomplished by Saha. However, before turning to Saha's work, let us take a brief look at the other provocative paper by Lindemann which also arrived at the ionization equation.

F.A. Lindemann [5] was interested in the problem of magnetic storms. These are sudden disturbances in the Earth's magnetic field. In 1859 Richard Carrington discovered an explosion on the Sun (a phenomenon we now call a 'solar flare') and came to know that there was a magnetic storm a few hours later. By the beginning of the twentieth century, there was enough circumstantial evidence that magnetic storms were caused by solar explosions. The problem was to work out the detailed mechanism of how this happens. Lindemann suggested the hypothesis that "an approximate equal number of positive and negative ions are projected from the sun in something of the form of a cloud and that these are the cause of magnetic storms and aurorae." In today's language, we would say that, according to Lindemann, plasma blobs would be thrown from the Sun during solar explosions. In order to substantiate this hypothesis, Lindemann had to show that the material in the solar atmosphere would be ionized. He used the same equation as Eggert (see Fig. 4) and showed that hydrogen would be ionized, commenting that hydrogen "is the only substance about which enough is known to enable quantitative results." Curiously, Lindemann does not cite either Nernst or Eggert, but merely states that the method he was using "is well known."

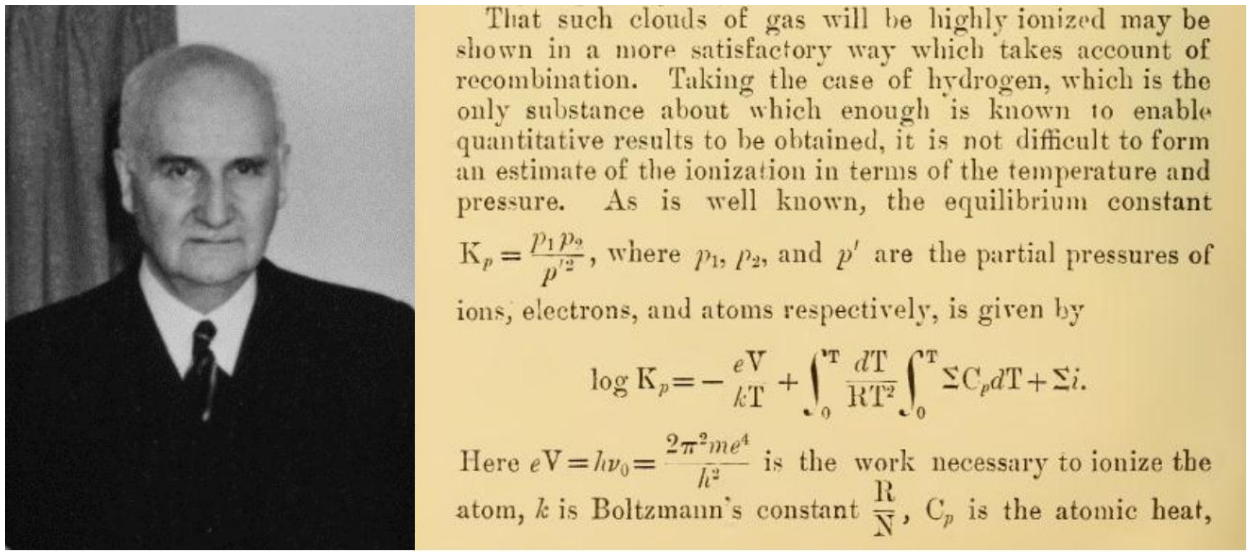

Fig. 4. Frederick Alexander Lindemann (1886 – 1957) and the part of his paper [5] where the ionization equation was written down.

Saha [21] had given credit to Eggert as the pioneer in his own paper and clearly mentioned that he took the crucial equation from Eggert's paper. However, Saha was not aware of Lindemann's work until he met Lindemann in 1921 at the Royal Astronomical Society when he (Saha) was visiting London and his first paper on thermal ionization had already been published. Saha wrote to Plaskett [17]: "I was presented by somebody, I do not remember now who it was, to Lindemann, who had read my paper on the Ionisation in the solar chromosphere, published in the February number of the Phil. Mag. He told me that he had published something of a similar type somewhat earlier in the Phil. Mag. I looked up the references given the next day and found that he had deduced the thermal ionization for hydrogen, taking the correct value of the I.P. for hydrogen." Lindemann, who worked at Oxford where Plaskett also worked, apparently felt bitter that he never got his due credit. Plaskett wrote to Saha [17]: "If there is any feeling of doubt about your work, it is probably to be found amongst the physicists in Oxford who feel that not sufficient credit is ever given to Lindemann's pioneer work in this same field." Plaskett, however, hastily added: "I do not share this doubt. It is true that Lindemann was the first to apply thermal ionization to the problem of the stellar atmosphere, but it was you who showed how fruitful this concept was in describing the stellar sequence in terms of the parameters of temperature and pressure."

Incidentally, coronal mass ejections (CMEs) were discovered more than half a century after Lindemann's work and triumphantly validated his hypothesis that plasma blobs thrown from the Sun during solar explosions cause magnetic storms. Although Lindemann's work should presumably be taken as the first theoretical prediction of CMEs, this work seems to be completely unknown to the present-day researchers in the field of CMEs and is not cited in the recent reviews on this subject [22].

Apart from anticipating Saha's work, Lindemann had another – rather unhappy – connection with India in his later life. As a scientific adviser to Winston Churchill during the Second World War, Lindemann (raised to peerage as Lord Cherwell) played a key role in the adoption of certain policies which led to the terrible Bengal famine of 1943 [23]. Saha himself was very much politically active at that time. I have often wondered if Saha knew about Lindemann's role in causing the Bengal famine and if they had any correspondence in later life. I have not been able to find any answer to these questions.

## Saha's own contributions

Now we come to Saha's own work. His epoch-making work on thermal ionization was presented in a series of four moderately long papers: "Ionisation in the solar chromosphere" [21], "Elements in the Sun" [24], "On the problems of temperature radiation of gases" [25] and "On a physical theory of stellar spectra" [26]. The first and the last papers were the most important. We shall now discuss what Saha achieved in these two papers, with some comments on the second paper. All these papers were prepared when Saha was a Lecturer at Calcutta University and were submitted to *Philosophical Magazine*. Almost immediately after the submission of these papers written in quick succession, Saha proceeded to England with a scholarship and worked in the laboratory of Alfred Fowler, an expert on spectroscopy. The first three papers appeared one after another in *Philosophical Magazine* (between October 1920 and February 1921) soon after Saha's arrival in England (the dates of submission of these papers from Calcutta were March 4, 1920; May 22, 1920; and May 25, 1920). However, the fourth and last paper in the series was eventually withdrawn from *Philosophical Magazine*, expanded in consultation with Fowler and was finally communicated by Fowler to *Proceedings of Royal Society*, in which it appeared. Saha wrote the following at the end of his last paper: "it is my pleasure to record my best thanks to Prof. A. Fowler for the interest he has taken in the work, and the many valuable items of information, advice and criticism with which he has helped me."

> The explanation of these problems, and some other associated problems of solar physics, will be attempted in this paper. The method is based upon a recent work of
>
> \* Kossel and Sommerfeld, *loc. cit.* p. 250.
> † This line is masked by the strong hydrogen line $H_\zeta$.
> ‡ Mitchell, *loc. cit.* pp. 490–491.
>
> 476          Dr. Megh Nad Saha *on Ionization*
>
> Eggert\*—"On the State of Dissociation in the Inside of fixed Stars." In this problem, Eggert has shown that by applying Nernst's formula of "Reaction-isobar,"
>
> $$K = \frac{p_M^{\nu_m} p_N^{\nu_n} \ldots}{p_A^{\nu_A} p_B^{\nu_B} \ldots},$$
>
> to the problems of gaseous equilibrium in the inside of stars, it is possible to substantiate many of the assumptions made by Eddington † in his beautiful theory of the constitution of stars. These assumptions are that in the inside of stars the temperature is of the range of $10^5$ to $10^6$ degrees and the pressure is about $10^7$ Atm., and the atoms are so highly ionized that the mean atomic weight is not much greater than 2. This method is directly applicable to the study of the problems sketched above. The equation of the Reaction-isobar is
>
> $$\log K = \log \frac{p_M^{\nu_m} p_N^{\nu_n} \ldots}{p_A^{\nu_A} p_B^{\nu_B} \ldots} = -\frac{U}{4 \cdot 571\, T} + \frac{\Sigma \nu C_p}{R} + \Sigma \nu C, \quad (1)$$
>
> where K = the Reaction-isobar,
>       U = heat of dissociation,
>       $C_p$ = specific heat at constant pressure,
>       C = Nernst's Chemical constant,
>
> and the summation is extended over all the reacting substances. The present case is treated as a sort of chemical reaction, in which we have to substitute ionization for chemical decomposition. The next section shows how U is to be calculated. The equation will be resumed in § 3.

Fig. 5. The part of the first paper on thermal ionization by Saha [21] where he refers to his predecessors.

Fig. 5 shows a portion from Saha's first paper on this subject where he acknowledges Eggert as his forerunner. As we have seen, Eggert [4] came very close to anticipating Saha, but failed to figure out how to obtain the heat of reaction for the ionization of any other substance besides hydrogen and did not realize that stellar spectra, rather than the stellar interior, happened to be the subject which awaited a major breakthrough on the application of the principles of thermal ionization. Saha's habit of reading "desultorily" led him to realize that he could get the values of the heat of reaction for many elements from the recent atomic physics experiments and he was also conversant with the problems of stellar spectroscopy. Saha [21] wrote the following about the heat of reaction: "The value of U in the case of alkaline earths, and many other elements, can easily be calculated from the value of the ionization potential of elements as determined by Franck and Hertz, MacLennan, and others." For any element for which ionization potential data were available, Saha could calculate the level of ionization and could apply the results to the study of a stellar atmosphere which was responsible for producing stellar spectra. Saha wrote the following on the basis of his personal interactions with Eggert in later years (Saha to Plaskett, 21 December 1947 [17]): "when I met him he told me that he had not realised the importance of Bohr's theory, or of Franck and Hertz's experiments, and had absolutely no idea that the ionisation potential can be calculated from spectroscopic data, as I had done. Further, he appears to have had no idea of problems of solar chromosphere or Physical characteristics of Stellar spectra, as was apparent to one like me who had read Miss Agnes Clerke's Book. His knowledge of the problem of the Stellar Interiors was derived from a talk given by Kohlochutter in the Berlin Physicale Colloquium on Eddington's Theories." A careful reading of Eggert's paper would convince anybody that Saha's assessment of Eggert was quite accurate. Certainly somebody trained in a conventional way in a narrowly specialized field of research could not do the synthetic work that Saha was to undertake. A maverick like Saha was needed for this work. We now turn to the problems of stellar spectroscopy which Saha solved by applying the theory of thermal ionization.

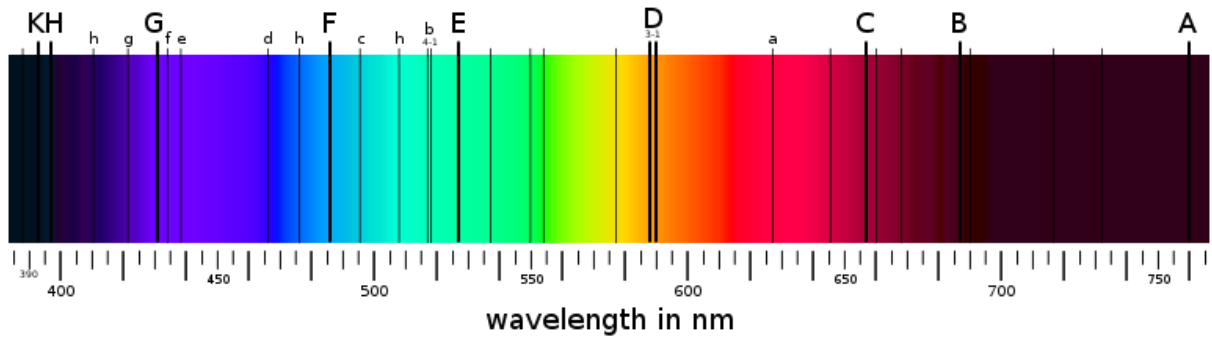

Fig. 6. Fraunhofer lines in the spectrum of sunlight.

J. Fraunhofer is credited with the 1817 discovery of many dark lines present in the spectrum of the Sun. R. Bunsen and G. Kirchhoff realized in 1861 that these are signatures of different chemical elements. Atoms of these elements present in the gaseous form in the solar atmosphere absorbed the passing radiation at certain frequencies. They also realized that these gases heated to a high temperature would emit in spectral lines at exactly the same frequencies. Now, what we see in the spectrum of the solar disk are absorption lines. However, if we can isolate light from the much less dense gas a little above the Sun's surface, we expect to see emission lines. Although the light from the gas above the solar surface is normally completely overshadowed by the much stronger light from the disk, there is one way of isolating such light from the gas above the solar surface. When the Moon covers the solar disk during a total solar eclipse, there is a very short interval of time during which we get radiation only from the gas above the solar surface. The spectrum of such radiation is called flash spectrum and is indeed found to consist of spectral lines. Since the light coming from the gas above the solar surface appears more colourful than normal sunlight, the layer of solar atmosphere from which this light comes is called the chromosphere ('chromo' is the Greek word for colour), shown in Fig. 7. One would naively expect that the emission lines of the chromosphere would be at the same frequencies as the absorption lines in the regular solar spectrum. But this was found not to be the case! The emission lines in the flash spectrum and the absorption lines in the regular spectrum of sunlight were often at very different frequencies. This was a completely baffling problem of stellar spectroscopy which Saha solved in his first paper [21].

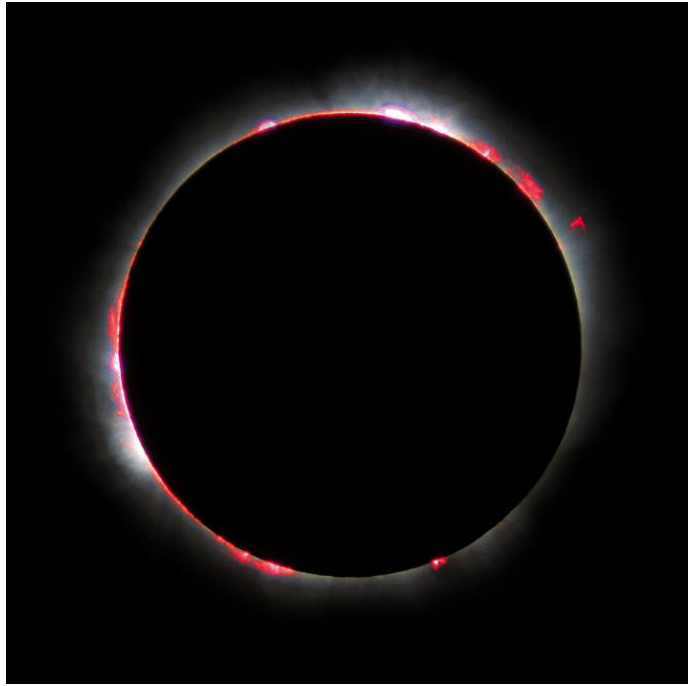

Fig. 7. Light coming from the chromosphere above the solar surface during a total solar eclipse. The spectrum of this light is called the flash spectrum.

The other problem of stellar spectroscopy which Saha solved in his last great paper of the series [26] was even more serious. Even a casual observation of stars readily shows that different stars are of different colour. The reddish stars presumably have lower surface temperature and bluish stars higher surface temperature. Astronomers found that stars of different colours had different kinds of spectral lines. Based on the spectral lines, a group of astronomers at the Harvard College Observatory, led by E.C. Pickering and A.J. Cannon, classified all the stars in a small number of spectral classes: O, B, A, F, G, K and M. Let us not get into the details of spectral classification. But do we have to conclude that reddish, yellowish and bluish stars have different chemical composition and that is why they have different spectral lines? This was undoubtedly the most outstanding question of stellar astrophysics when Saha started working on the theory of thermal ionization.

Although quantum mechanics was yet waiting to be born, the Bohr-Sommerfeld model available at that that time gave an account of the atomic structure and the spectral lines of the atom on the basis of the old quantum theory. According to this model, atoms with a small number of electrons in their outermost shells are expected to have spectral lines at reasonably regular frequencies. With astute intuition, Saha realized that alkaline earth elements like calcium would be particularly useful for his studies. A normal atom of calcium has two electrons in its outermost shell, whereas a singly ionized calcium atom (denoted as $Ca_+$) has one electron in its outermost shell (like a hydrogen atom). So, alkaline earth atoms have a small number of electrons in their outermost shell both before and after ionization. As a result, both the neutral spectrum and the ionized spectrum are expected to have regular patterns. Physicists knew that the spectrum obtained in a spark was somewhat different from the normal spectrum of an element. Norman Lockyer, who studied this problem, was under the impression that atoms under the higher temperature of the spark emit 'enhanced lines' [27]. But, in Lockyer's time, when atoms were still viewed as indivisible entities of matter, it was completely unclear why atoms would behave so differently on increasing the temperature. Saha gave a simple argument why the spark spectrum should be due to ionized atoms. It was known that the spark spectrum of calcium has spectral lines at frequencies approximately four times the frequencies of normal calcium. Saha realized this to imply that the effective Rydberg constant of calcium atoms producing the spark spectrum to be four times that of normal calcium atoms. Now, according to the Bohr model of the atom, the Rydberg constant goes as the square of the charge seen by the electron taking part in the production of the spectral line. Now, the outermost electron in a $Ca_+$ atom would see an effective charge of $2e$, because the net charge of the nucleus plus the inner shells is $2e$. This would make the effective Rydberg constant for $Ca_+$ to have four times its normal value. From such considerations, Saha argued that the atoms of calcium producing the spark spectrum must be singly ionized $Ca_+$ atoms. In other words, the spark spectrum of an element is the spectrum produced by its ionized form.

It was known that the regular spectrum of sunlight has an absorption line of normal calcium called g line (at wavelength 422.7 nm) and also absorption lines of the spark spectrum of calcium (i.e. inonized calcium $Ca_+$) called H and K lines (at wavelengths 396.9 and 393.4 nm). All these lines can be seen in the Fraunhofer spectrum of the Sun shown in Fig. 6. However, the flash spectrum for the upper chromosphere showed no g line and much stronger H and K lines. Saha [21] suggested that "The high-level chromosphere is … the seat of very intense ionization". This would explain the absence of the g line of normal calcium and the high strength of the spark lines of ionized $Ca_+$ in the flash spectrum from the chromosphere. We now know that the temperature increases as we go up in the solar atmosphere, for reasons we shall not discuss here (see [28], Chapter 8). But this was not known in 1920. It was thought the chromosphere is colder than the solar surface and one naively expected a lower ionization in the chromosphere because of the lower temperature. On the basis of the ionization equation, Saha showed that the level of ionization depended crucially on pressure apart from temperature. Fig. 8 from Saha's paper [21] shows a table giving the percentage ionization of calcium computed by Saha from the ionization equation. As the pressure falls with height in the higher solar atmosphere, Saha's calculations showed that ionization of calcium would increase with height in the chromosphere. In the high chromosphere, there would be no neutral calcium atoms left to produce the g line, but many Ca+ atoms to give rise to H and K lines. Saha did similar calculations for other alkaline earths (such as strontium and barium) as well and suddenly there was an explanation of the broad features of the flash spectrum.

TABLE IV.

Ionization of Calcium (in per cents.).
U = 6·12 volts = 1·40 . 10⁵ calories approximately.
Pressure in atmospheres—Temperature on the Absolute Scale.

| Pressure ... | 10. | 1. | $10^{-1}$. | $10^{-2}$. | $10^{-3}$. | $10^{-4}$. | $10^{-6}$. | $10^{-8}$. |
|---|---|---|---|---|---|---|---|---|
| Temp. | | | | | | | | |
| 2000° ...... | | | | | | $5 . 10^{-4}$ | $1·4 . 10^{-3}$ | |
| 2500 ...... | | | | | | $2 . 10^{-2}$ | $7 . 10^{-2}$ | |
| 3000 ...... | | | | | | $3 . 10^{-1}$ | 1 | 9 |
| 4000 ...... | | | | | 2·8 | 9 | 26 | 93 |
| 5000 ...... | | | 2 | 6 | 20 | 55 | 90 | |
| 6000 ...... | | 2 | 8 | 26 | 64 | 93 | 99 | |
| 7000 ...... | | 7 | 23 | 68 | 91 | 99 | | |
| 7500 ...... | | 11 | 34 | 75 | 96·5 | | | |
| 8000 ...... | | 16 | 46 | 84 | 98·5 | | | |
| 9000 ...... | | 29 | 70 | 95 | | Complete | | |
| 10000 ...... | | 46 | 85 | 98·5 | | Ionization. | | |
| 11000 ...... | | 63 | 93 | | | | | |
| 12000 ...... | | 76 | 96·5 | | | | | |
| 13000 ...... | | 84 | 98·5 | | | | | |
| 14000 ...... | | 90 | | | | | | |

Fig. 8. A table giving the percentage ionization of calcium at different temperatures and pressures. From the first paper on thermal ionization by Saha [21].

After considering the alkaline earths in the first paper [21], Saha turned his attention to the alkalis in the second paper [24]. He addressed the question of why the lines of alkalis like potassium, rubidium and caesium are either absent in the solar spectrum or are present very weakly. On applying the ionization equation, Saha found that these alkali atoms would be almost fully ionized under the conditions of the solar atmosphere. A singly ionized alkali atom has a full outermost shell and usually does not produce lines in the visible part of the spectrum. So we do not see lines from most of the alkali elements in the solar spectrum. However, Saha estimated that, at the somewhat lower temperatures prevailing inside sunspots, many of these alkalis would be only partially ionized. As a result, Saha argued that spectral lines of normal alkali atoms should be much stronger in the spectra of sunspots. This was known to be the case for the $D_1$ and $D_2$ lines of sodium. Saha predicted that spectral lines of potassium, rubidium and caesium (absent in normal Fraunhofer spectrum from the solar disk) should also be visible in the spectra of sunspots and asked astronomers to look for them. We shall discuss later how astronomers followed up on this prediction.

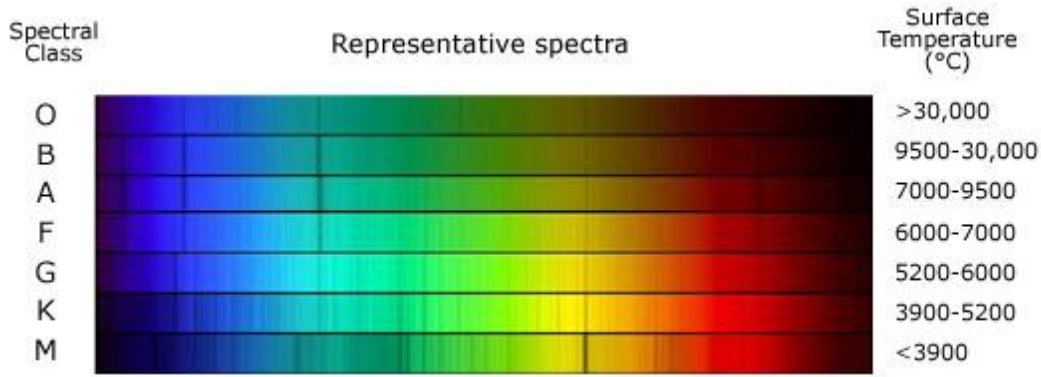

Fig. 9. Spectra of stars of different classes stacked above one another. M stars (reddish in colour) have the lowest surface temperature, whereas O stars (bluish) have the highest. Our Sun belongs to the class G.

In the last of the four great papers [26], Saha finally turned his attention to one of the most outstanding problems of astrophysics of his day – explaining the spectral sequence of stars. It was already realized at that time that the various spectral types (O, B, A, F, G, K, M) formed a continuous sequence, there being many stars lying between two types. When Saha entered the field, it was already guessed that this sequence might arise from the variation of some physical parameter like the surface temperature of the stars. It was not clear why spectral lines of different elements were present in the different spectral classes. Many astronomers thought that this may imply different compositions. Saha managed to put all the pieces of the jigsaw puzzle together in a single stroke with his theory of thermal ionization. Let us consider the case of calcium. In the spectra of M stars (which are of reddish colour and are presumably cooler than other stars), it was found that the g line of normal calcium is very strong, but H and K lines of $Ca^+$ are barely visible. As we move along the spectral sequence towards O stars (bluish and presumably hotter), the g line disappears completely by the time we reach B stars, whereas H and K lines become very strong for A stars. As we move further towards O stars, the H and K lines start becoming weaker and eventually disappear completely in the very hot O stars. In the very first paper on thermal ionization, Saha had already calculated the ionization of calcium by using his ionization formula for different temperatures (Fig. 8). Assuming the pressure to be one atmosphere (the typical pressure expected in a stellar atmosphere), Saha found that calcium becomes completely ionized to $Ca_+$ around temperature 13,000 K (see Fig. 8). Since the g line of calcium is not seen in stars hotter than B stars, he concluded that the surface temperature of those B stars in which the g line is barely seen must be about 13,000 K. On increasing the temperature further, Saha estimated that $Ca_+$ would be ionized further to doubly ionized $Ca_{++}$, this second-step ionization being nearly complete at about 20,000 K. His argument was that, for those stars in which the H and K lines of $Ca_+$ disappear, the surface temperature should be like this. Considering appearances and disappearances of many spectral lines in different spectral classes, Saha succeeded in mapping the entire spectral classification to a temperature scale.

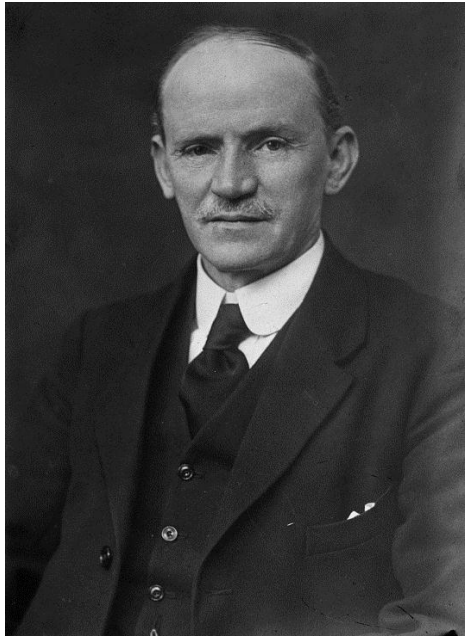

Fig. 10. Professor Alfred Fowler (1868 – 1940) of the Imperial College, London. Saha revised the last of his four great papers on thermal ionization while he was a visitor in Fowler's laboratory.

At last the mystery of why stars of different colours had different kinds of spectra was completely solved. Stars of different colours do not have different compositions! Rather, they have different surface temperatures at which atoms of various elements would be ionized to different extents, causing the resulting spectral lines to be different. Saha triumphantly ended the paper [26] with the declaration that "the theory is only a first attempt for quantitatively estimating the physical processes taking place at high temperature. We have practically no laboratory data to guide us, but the stellar spectra may be regarded as unfolding to us, in an unbroken sequence, the physical processes succeeding each other as the temperature is continually varied from 3000 K to 40,000 K."

It may be mentioned that the original version of Saha's fourth great paper was titled "On the Harvard classification of stellar spectra", as mentioned in a synopsis of his four papers on thermal ionization which Saha prepared on the eve of his departure to Europe at the completion of these papers [29]. Since Pickering and Cannon worked at the Harvard College Observatory, the stellar classification scheme which they developed is often referred to as the Harvard classification. When Saha was revising this paper during his stay in Alfred Fowler's laboratory, he changed the title of the paper to "On a physical theory of stellar spectra" at the suggestion of Fowler, who felt that the earlier title did not reflect the fact that "pioneering credit for such researches must be given to Lockyer", who was Fowler's mentor. Since no copy of Saha's original paper is known to exist, we cannot assess the amount of change which Saha made while revising the paper. Saha's own account of it is as follows (Saha to Plaskett, 21 December 1946, [17]): "I took about four months in rewriting this paper, and all the time I had the advantage of Prof. Fowler's criticism, and access to his unrivalled stock of knowledge of spectroscopy and astrophysics. Though the main ideas and working of the paper remained unchanged, the substance matter was greatly improved on account of Fowler's kindness in placing at my disposal fresh data, and offering criticism wherever I went a little astray, out of mere enthusiasm."

A close reading of Saha's fundamental papers makes it clear that he was not an 'expert' in any of the diverse fields which he was combining in a grand synthesis. He repeatedly used 7500 K as the temperature of the solar surface, whereas the actual temperature is close to 5800 K. In a letter to H.N. Russell who carried out extensive analysis based on Saha's theory, another astronomer S.A. Mitchell commented (quoted in [10]): "Apparently Saha is not very well up on matters of Astronomical interest, and being a physicist we could not expect him to be. His value of 7500 K for the Sun was very surprising." This shows the extreme isolation in which Saha did his great work. While he was working in Calcutta, there was no competent astronomer around him with whom he could check the surface temperature of the Sun.

## The aftermath of Saha's great work

Truly great works of science often have a compelling simplicity. Many of Saha's ideas may seem simple and obvious to us today. It may even be difficult to imagine the impact this work made on the contemporary astrophysics scene. Several leading astrophysicists of England and America almost immediately started important investigations following on Saha's lead. To give an indication of the impact of Saha's work, Figs. 11 and 12 show the beginnings of important contemporary papers by well-known authors [30–31] who start by referring to Saha's work. In the textbook *Theoretical Astrophysics* published in 1936, the well-known astrophysicist S. Rosseland wrote [32]: "The impetus given to astrophysics by Saha's work can scarcely be over-estimated, as nearly all later progress in this field has been influenced by it, and much of the subsequent work has the character of refinements of Saha's ideas."

There were several extensions of Saha's work. The ionization of an element, say calcium, will depend not only on the electrons which had come out of the calcium atoms (as Saha assumed), but on all the electrons which had come out of different atoms. Russell [30] developed the theory to address this issue. In a letter to Russell dated 26$^{th}$ April 1922, Saha wrote [15]: "I have been quite convinced of the correctness and importance of your extension of the theory to the case where several gaseous mixtures are present. My first impression was that the extension was of an "ad hoc" nature, but I find that I was mistaken." In Saha's original theory, the various possible excited states of an atom at a particular ionization level were not taken into account. This extension was done by Ralph Fowler [33], who was later to become the doctoral thesis supervisor of two famous physicists of Indian origin – S. Chandrasekhar and H.J. Bhabha. Afterwards, Fowler and Milne [31] showed that the application of Saha's theory was simpler if one considered the stellar spectral class in which a particular line had the maximum strength instead of considering the spectral classes in which the line appeared and disappeared (as Saha had done).

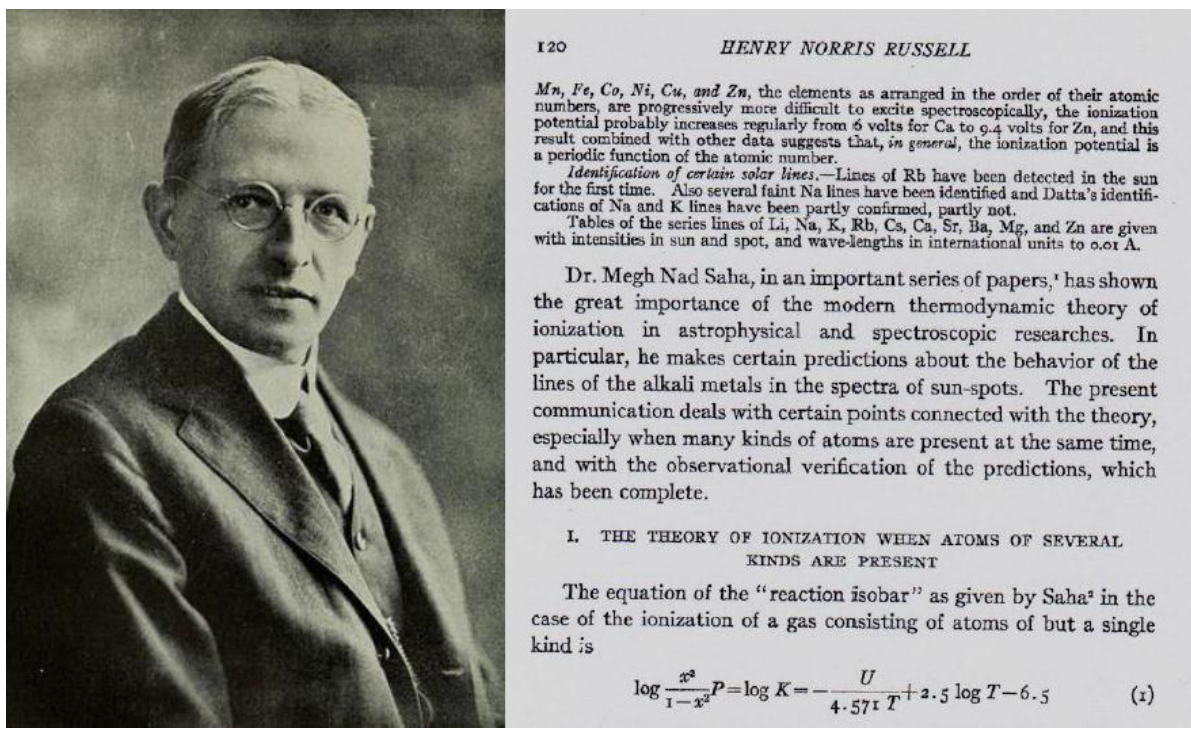

Fig. 11. Henry Norris Russell (1877–1957) and the beginning of his paper [30] following up on Saha's work.

What was Saha himself doing? When the fourth and the last of his great papers appeared, he was still less than 28. It is unlikely that a brilliant young man of 27 would think that his best work was already behind him. After all, Saha had just got a foothold in the scientific community after his initial years of struggle. Probably he thought that his work on thermal ionization was just the beginning of an illustrious scientific career that would scale still greater heights. Unfortunately, that was not to be. Saha had planned two kinds of follow-up works after his four great papers. Firstly, he wanted to carry on laboratory experiments on thermal ionization. Secondly, he wanted to follow up on what he had predicted in his second great paper, that the spectral lines of alkali metals would be stronger in the spectra of sunspots compared to the spectra of normal solar surface. However, Saha could not become a key player in either of these ventures. As I have repeatedly emphasized, Saha had some advantages as an outsider when he was doing his great synthetic work. A young person trained conventionally for research in a narrow area of science would not have the knowledge of diverse unrelated areas of science which Saha had acquired through his habit of reading "desultorily". But, once that great synthetic work requiring such knowledge was over, Saha immediately lost this special advantage that he had. The follow-up works could be done best by more narrowly focused professional scientists working in well-equipped places. Saha quickly realized that he could no longer remain in the cutting-edge research frontier.

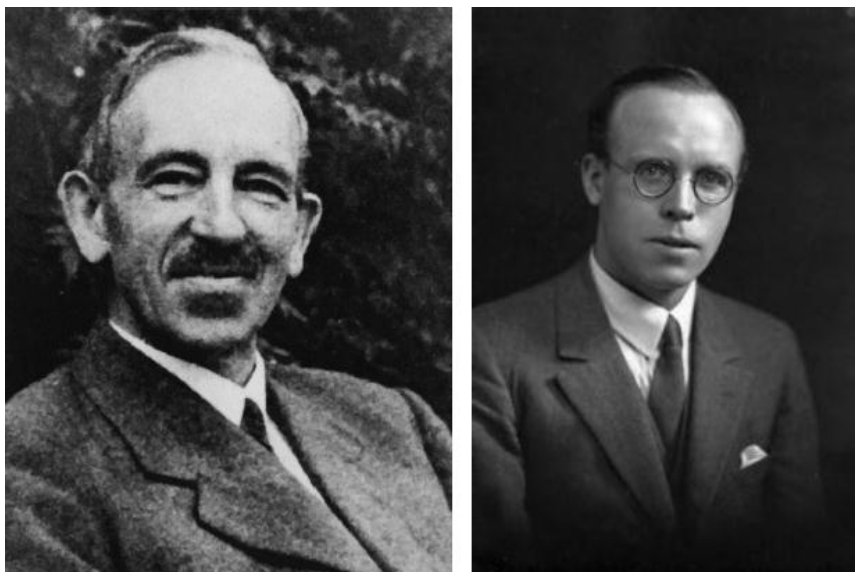

Fig. 12. Top left: Ralph Howard Fowler (1989–1944), and top right: Edward Arthur Milne (1896–1950), with the beginning of their paper [31] following up on Saha's work.

After spending four months (from November 1920 to February 1921) at the Imperial College in Fowler's laboratory, Saha decided to move to Nernst's laboratory in Berlin, which had better facilities for carrying on experimental work on thermal ionization than any place in England. According to Saha (Saha to Plaskett, 21 December 1946 [17]): "I had written to Eggert who was Nernst's assistant, and asked him to get Nernst's permission for me to work at his laboratory. I received an encouraging reply both from Eggert and Nernst himself and went to Berlin in February, 1921." In fact, Fowler had wanted Saha to continue in his laboratory "as he thought that spectroscopic part of the work regarding stellar spectra required to be worked out in greater detail". However, when Fowler realized that Saha was really bent upon doing experimental work at Nernst's laboratory, "Fowler agreed rather ruefully." About his experience in Berlin, Saha wrote: "On arrival at Berlin, Nernst received me very warmly, and gave me facilities at his laboratory for verifying the theory experimentally." But this bonhomie was short-lived. Saha's experimental work at Nernst's laboratory did not go well and did not result in any published paper. In a letter to Vice-Chancellor Asutosh Mookerjee of Calcutta University dated 20-8-21, Saha had proudly boasted [15]: "I have succeeded in experimentally demonstrating that gaseous atoms can be ionised simply by heat (temperature ionization of gases) – an effect which under various forms have been looked for in vain by such eminent physicists as Maxwell, Hittorf, Rayleigh and J.J. Thomson. The experiment will shortly be published in the Zeitschrift für Physik, and either myself or Prof. Nernst will announce it before the forthcoming meeting of the German physicists at Jena, where I also have an invitation." This paper promised to Mookerjee never materialized! Could this be due to the lack of experimental skill on the part of Saha? Both C.V. Raman and S.N. Bose thought that Saha was not good at experiments. After Saha returned to India and shifted to Allahabad University, he tried to set up experimental facilities there for studying thermal ionization. Initially he had difficulties in obtaining funds. However, eventually he managed to get some funds from the Governor of the province when he was elected FRS. This enabled Saha to start his experiment on thermal ionization. But the outcome was not spectacular. It finally led to one short paper in Zeitschrift für Physik written by Saha with his students [34]. Fig. 13 shows the furnace which Saha constructed at Allahabad for this experimental work.

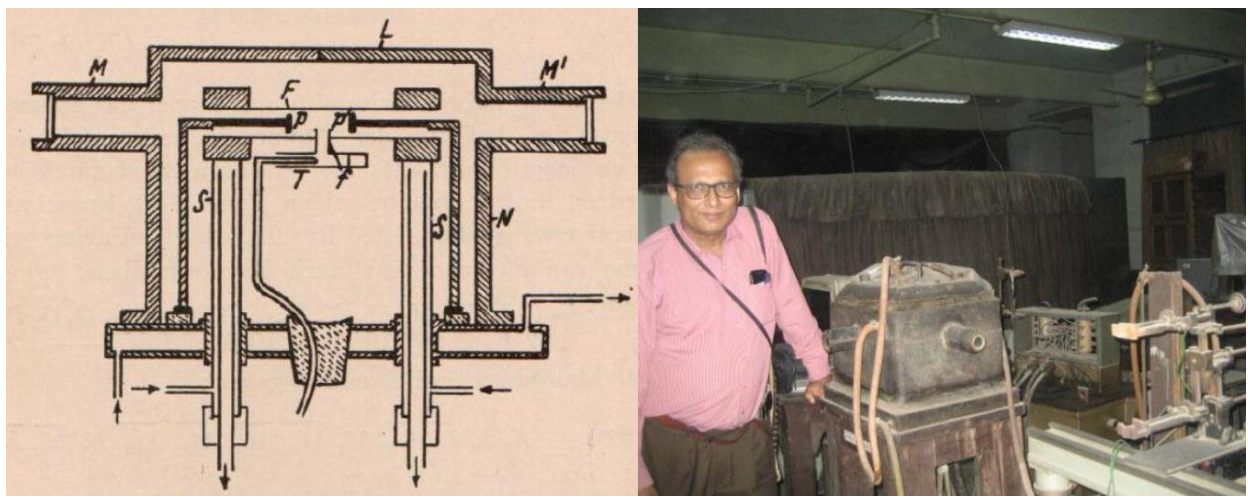

Fig. 13. The left panel shows the design of the furnace used in the experimental studies of thermal ionization, as given in the paper by Saha, Sur and Mazumdar [34]. This furnace is still kept in the basement laboratory of the Physics Department of Allahabad University. The right panel shows the author standing next to the furnace.

Let us now turn to Saha's prediction for the existence of absorption lines of alkalis in the spectra of sunspots [24]. The Mount Wilson Observatory near Los Angeles had the best solar observing facilities in the world to check this. On 9 July 1921, Saha wrote a letter from Berlin to the great solar physicist George Ellery Hale, who was the Director of this Observatory. After introducing himself as the author of the recent papers on thermal ionization, Saha wrote [15]: "I believe that these theories explain a good deal of the valuable data accumulated during the last 40 years by the American and English astrophysicists, and besides open up new paths of investigation. I append herewith an account of some investigations which may immediately be taken up. . . I shall be very glad if some one at the Mount Wilson Solar Observatory undertakes the works suggested overleaf. My means are too limited, and as my University is very poorly provided for astrophysical work, I see no prospects of ever being able to carry out the ideas contained in my papers." Saha probably did not know that Hale had become a serious patient of Schizophrenia by that time. With deteriorating health condition, Hale had to resign from the Directorship of the Observatory around that time.

Luckily, Saha's letter fell in the hands of the great Princeton astrophysicist H.N. Russell, who was a research associate at the Mount Wilson Observatory at that time and undertook the work suggested by Saha. In a letter dated August 3, 1921 (less than a month from the date of Saha's letter!), Russell wrote to Saha [15]: "Dr. Hale has shown me your recent letter to him. . . your predictions about the lines of alkali metals have been completely verified. Rubidium is present. We will investigate caesium as soon as proper photographs of the spot spectrum near $\lambda$ 8500 can be obtained." Fig. 11 shows the beginning of Russell's great 25-page paper in *Astrophysical Journal* presenting his investigations on this subject. Interestingly, towards the end of the abstract of this paper, Russell mentions "Datta's identifications of Na and K lines" (can be seen in Fig. 11). This was Saha's college batchmate Snehamoy Datta (he was a student in the Physics Department of Presidency College in Calcutta, whereas Saha was in the Mixed Mathematics Department). Datta was working in Alfred Fowler's laboratory in London when Saha arrived in England and urged Saha also join this laboratory. After the classic paper shown in Fig. 11, Russell wrote several other papers on verifying Saha's theory, heaping generous praise on Saha for his pioneering theoretical work. Probably these papers produced a mixed reaction in Saha. On the one hand, he must have been happy that his predictions were being shown to be correct and he was being given full credit for making those predictions. On the other hand, Saha must have realized that the research field which he had initiated had already gone out of his hands and probably he would never be a major player in that research field again.

Since Saha was in Europe when his pioneering papers were coming out, one might have thought that he was closer to the scene of action. However, the supreme irony is that probably Saha could himself participate in the follow-up works had he been in India! The second best place in the world at that time for the type of solar observational work that was undertaken at the Mount Wilson Observatory was the Kodaikanal Observatory in south India. It was headed by a very capable astronomer John Evershed, who had developed state-of-the-art spectroscopic observational facilities there (see Fig. 14) and in 1909 had discovered the famous Evershed effect

– a type of flow pattern around sunspots – from the spectroscopic study of sunspots [35]. Unfortunately, Evershed, although competent in instrumentation and astronomical observations, did not have much understanding of theory. In a letter to M. Minnaert, Saha disclosed a little-known fact (Saha to Minnaert, October 19, 1955, [15]): "When I wrote my first paper on "Thermal Ionisation" published in Phil. Mag. Vol 39, 1920, I asked Evershed to communicate it to the M.N.R.A.S. He refused saying that he could not understand the paper. I sent it straight to the M.N.R.A.S.; it was refused forthwith." (Here M.N.R.A.S. stands for *Monthly Notices of Royal Astronomical Society*, the leading British journal in astronomy). However, Evershed must have realized his oversight soon. When Saha was nominated for the Fellowship of the Royal Society in 1925, Evershed was one of the signatories in the nomination certificate [36]. I have never come across any account of a personal meeting between Evershed and Saha. I do not know if they ever met each other. When Saha wrote to Hale, I wonder why he did not explore if observations could be done at Kodaikanal to verify his theoretical predictions. Was Saha still feeling bitter that Evershed had not communicated his paper when he was still an unknown young scientist? Had Saha been in India, presumably he himself could go to Kodaikanal and carry on this work of verifying his theoretical predictions. By the time Saha returned to India, all the important observational studies had already been done at the Mount Wilson Observatory at a very fast pace and Evershed was getting ready to leave for England on his retirement in 1923.

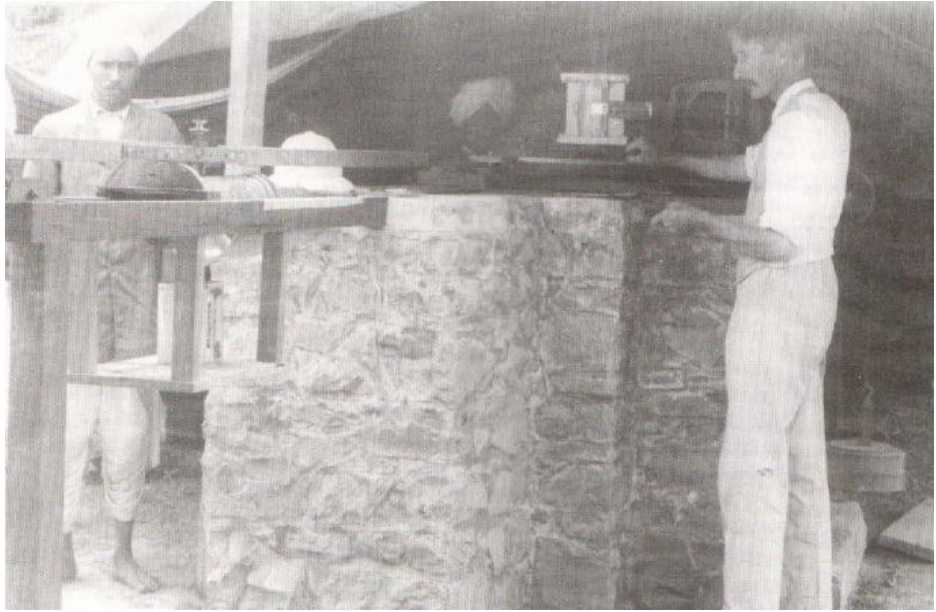

Fig. 14. John Evershed (1864–1956) working with the spectroheliograph at the Kodaikanal Observatory. Saha's theoretical predictions probably could have been verified with this instrument attached to the back-end of the solar telescope.

With growing international reputation, Saha started receiving various recognitions. Alfred Fowler took the initiative of nominating Saha for the Fellowship of the Royal Society in 1925.

Saha was elected in 1927 – being only the fourth Indian to become FRS in the early decades of the twentieth century, after S. Ramanujan, J.C. Bose and C.V. Raman [37]. Saha wrote in his letter to Plaskett [17]: "I think I owe my election in 1927 chiefly to him [Fowler], Eddington and Chapman." Curiously, Eddington and Chapman were not signatories in Saha's nomination certificate [36]. One signatory was Lindemann, although he had some hard feeling that he did not get enough credit for his work on thermal ionization before Saha. It was known that Saha in his youth had close associations with the revolutionaries fighting the British empire. Before electing Saha, the Royal Society made some discreet enquiries whether this background of Saha might prove an embarrassment for the Society if Saha was elected. An account of this has been given by DeVorkin [7]. Saha was also nominated for the Nobel Prize. The Nobel Prize had never been given to an astrophysicist in the early part of the twentieth century, until Bethe won it in 1967 for "discoveries concerning energy production in stars". So Saha had very little chance. Interestingly, he was shortlisted and considered seriously in the year 1930 in which Raman eventually won the Prize. Discussions on Saha's nomination within the Nobel Prize Committee have been uncovered from the archives of the Nobel Prize by Singh and Riess [38].

When Saha left for England, he was a Lecturer at Calcutta University. As the news of the international appreciation of Saha's work reached the Calcutta academic circles, he was selected for the newly established Khaira Professorship of physics. Saha had to return from Europe to join in this position. However, he soon left for a professorship at Allahabad University. As I have already pointed out, there he tried to develop an experimental setup for the study of thermal ionization in the laboratory. However, Saha was never again to reach the height which he had reached at a very early age with his theory of thermal ionization. Saha again returned to Calcutta in 1937 as the Palit Professor at Calcutta University and became increasingly involved with science organization. His role in this phase has been studied by Anderson [39].

In the 1940s, Saha again turned his attention to the Sun, which was the subject of his first famous paper on thermal ionization. By that time, it was known that the temperature of the solar corona is much higher than that of the solar surface – a discovery in which the theory of thermal ionization played a crucial role. A red emission line and a green emission line of the corona were identified as coming from iron atoms which had lost 9 and 13 electrons respectively. Such a high degree of ionization required temperatures of the order of million degrees. Saha wrote embarrassingly wrong papers proposing that the high temperature of the corona was due to nuclear fission reactions taking place at the solar surface and in the corona [39]. A historian of science, of course, should not look at a work of science from the present-day vantage point. However, on the basis of the physics knowledge available at that time, Saha's theory appeared unlikely to most of the contemporary scientists. The astrophysicist M. Minnaert pointed out in a letter that Saha was plainly wrong (Minnaert to Saha, 30 November 1955 [15]): "about the corona I cannot well assume that you are right. . . the hypothetical nuclear process seems to me very unsatisfactory; the atomic velocities needed for such process would be much higher than our one million degrees!" Obviously, Saha had lost his magic touch. Reading Saha's papers on the solar corona, one gets the feeling that contemporary astrophysics was slipping through the fingers of the man who was hailed as the father of modern astrophysics just a few years earlier.

Just as Saha's appearance in the world of astrophysics was sudden, his disappearance as a major contributor to the science of astrophysics was also dramatic.  It almost seemed that his appearance and disappearance were like the appearance and disappearance of an absorption line along the sequence of stellar spectra! It must have been a matter of wonder to Saha's admirers within the international astrophysics community why the brilliant phase of Saha's creativity lasted for such a short time.  After reading Saha's account of his life in his letter to Plaskett, Russell wrote to the fellow astrophysicist Harlow Shapley, another admirer of Saha, in a letter dated Februay 13, 1947 [15]: "I had often wondered why Saha's activity in the theoretical field fell off after his return to India, and his account gives the explanation. . . his recent contributions to theory do not seem to me to be at all of the same importance as the old.  His history affords a very good argument that the Indian universities and perhaps the government should be actively interested in securing proper opportunities for men of proven ability."

As the international astrophysics community felt perplexed about Saha's sudden appearance on the scene with his early brilliant papers, doubts were often expressed whether Saha could have done these works in India completely on his own. Since Saha's papers appeared after he started working at Fowler's laboratory, it was widely believed that Fowler must have been some kind of supervisor for Saha.  Saha was aware of this and was intensely sensitive on this matter.  Plaskett had written in an article in *The Observatory*, April 1946 [17]: "Saha, working in Fowler's laboratory after the end of the last War, then demonstrated that the successive appearances of these different spectra could be interpreted as being due to the temperature and pressure prevailing in the stellar atmosphere." Although this may appear like an innocuous statement to us, this was enough to goad Saha into writing a long letter to Plaskett giving a kind of scientific autobiography, from which I have quoted extensively.  Saha wrote to Plaskett [17]: "I regret to tell you that this remark is entirely gratuitous and misleading.  It gives one the impression that I derived all the fundamental ideas in astrophysics which goes under my name viz., The Theory of Thermal Ionisation and Its Applications, from Prof. A. Fowler, while I was working as a scholar under him in 1920.  This is entirely misleading, for I worked at Fowler's laboratory as a guest . . ." Then Saha went on to point out that his first three papers on thermal ionization were written before he stepped out of India and only the initial version of the fourth paper was revised while Saha was at Fowler's laboratory.  Finally Saha summed up: "Fowler treated me as a colleague and a guest, never as a student. I have very great respect for him as a scholar, as a man, as an astrophysicist, and for the unselfish and generous way he treated me, which I now find is rather unique, and I am quite sure that were he living now, he would have been the first man to resent your suggestion."  On receiving Saha's letter, Plaskett replied immediately (Plaskett to Saha, 6 January 1947 [17]): "What was quite new to me was the fact that the early part of your work was done in India, not Germany, before you came to Fowler's laboratory.  The knowledge that you had done so much without help and backing in India only serves to increase the admiration I have always felt for your great contribution to astrophysics." It appears that many astrophysicists saw Saha's letter to Plaskett. Although Russell had corresponded extensively with Saha, he was also very much surprised by this letter. Russell wrote to Shapley (Russell to Harlow Shapley, February 13, 1947 [17]): "I do not think I knew till seeing this letter how much he had done previously in India."

## Conclusion

In my own textbook on astrophysics [41] published in 2010, I had written: "Saha (1920) derived the equation which tells us what fraction of a gas will be ionized at a certain temperature *T* and pressure *P*." I had not yet made a study of Saha's original papers at the time of writing this. However, such statements can be found in many textbooks including the one by Fermi [1]. Saha took the equation existing in the literature of thermal physics, figured out how to do quantitative calculations with it for elements other than hydrogen by using data taken from atomic physics experiments and provided the theoretical framework to understand stellar spectroscopy – a subject that had baffled everyone till Saha could put everything in order. Since Saha did not derive the basic equation or was not the first person to arrive at it, was it justified that the equation got named after Saha? How names get attached to important discoveries in science is a complex and not always a fair process. The fact that is undisputed is that Saha was the first person to realize that this equation, when combined with data from atomic physics, could unlock the mystery of stellar spectra. It is this application of the thermal ionization equation that immediately drew the attention of the entire astrophysics community by solving several long-standing puzzles of astrophysics and by making clear new predictions which could be verified quickly. It was only natural that Saha's papers would be the starting point from which all further follow-up works in this subject would take off and Saha's name would get associated with the whole subject. While pioneers like Eggert or Lindemann might have felt that they did not receive enough credit, I have not come across any statements in the literature of the subject which question the appropriateness of naming the equation of thermal ionization after Saha.

The aim of this article had been to analyze the specific nature of Saha's scientific contributions. Young people get trained in cutting-edge frontiers of science at established centres, and peripheral places very much outside such centres usually do not have the right kind of intellectual atmosphere to foster scientific creativity [42]. However, I have argued here that a brilliant person in a peripheral place without a proper training may occasionally have an advantage. Through his habit of reading "desultorily", Saha acquired a knowledge of many unrelated fields which was necessary for the paradigm-shifting synthetic work that he undertook. A person professionally trained in a more focused manner at an established centre of science was hardly expected to have the unusual technical background necessary to embark on such a synthesis. However, once the great synthesis was over, the time came for more routine research to tie up the loose ends. Saha was completely unable to compete in this game and was easily beaten by others working in established places with access to better facilities. Saha's case as a typical case of a peripheral scientist has also been analyzed by Dasgupta [8]. Saha's tale of extraordinary scientific achievements is simultaneously a tale of triumph and defeat, a tale both uplifting and tragic. Saha showed what a man coming from a humble background in an impoverished colony far from the active centres of science could achieve by the sheer intellectual power of his mind. But his inability to follow the trail which he himself had blazed makes it clear that there are limits to what even an exceptionally brilliant person could achieve in science under very adverse circumstances.

**Acknowledgments.** I thank Enakshi Chatterjee, D.C.V. Mallik, Atri Mukhopadhyay, Chitra Roy and Rajinder Singh for helping me to track down various source materials used in my research on Saha. Discussions with Deepanwita Dasgupta were particularly helpful in formulating some ideas presented in this paper. My research is supported by a JC Bose Fellowship awarded by DST.

**References:**

[1] E. Fermi, *Thermodynamics*, Dover, 1956, pp. 151–154.
[2] L.D. Landau and E.M. Lifshitz, *Statistical Physics*, 3rd edn., Part I, Pergamon, 1980, pp. 313–314.
[3] F. Reif, *Statistical and Thermal Physics*, McGraw-Hill, 1965, pp. 363–365.
[4] J. Eggert, Über den Dissoziationszustand der Fixsterngase, *Physik. Zeitschr.* **20** (1919) 570.
[5] F.A. Lindemann, Note on the Theory of Magnetic Storms, *Phil. Mag.* **38** (1919) 669.
[6] G. Venkataraman, *Saha and His Formula*, Universities Press, 1995.
[7] D.H. DeVorkin, Saha's Influence in the West: A Preliminary Account, in S.B. Karmohapatra (ed.) *Meghnad Saha Birth Centenary Commemoration Volume*, Saha Institute of Nuclear Physics, 1993.
[8] D. Dasgupta, Stars, Peripheral Scientists, and Equations: The Case of M. N. Saha, *Physics in Perspective* **17** (2015) 83.
[9] S. Dattagupta, On the Saha Ionization Equation, *Resonance* (January, 2018) 41.
[10] A. Mukhopadhyay, From Atoms to Stars: Meghnad Saha (1893-1956), *Ind. J. of Hist. Sci.* (2018) in press.
[11] S.N. Sen (ed.), *Professor Meghnad Saha: His Life, Work and Philosophy*, Meghnad Saha Sixtieth Birthday Committee, 1954.
[12] S. Chatterjee and E. Chatterjee, *Meghnad Saha*, National Book Trust, 1984.
[13] A. Mukhopadhyay, *Abinash Meghnad Saha* (in Bengali), Anushtup, 2012.
[14] A.R. Choudhuri, The Golden Age of Calcutta Physics: Difficulties in Reconstructing the History, in CU Physics 100, 2016, p. 1.
[15] M.N. Saha Archives, Saha Institute of Nuclear Physics.
[16] S.N. Bose, Amar Vijnan Charchar Purakhanda (in Bengali), in *Satyendra Nath Basu Rachana Sankalan*, Bangiya Vijnan Parishad, 1990, pp. 223–230.

[17] Saha-Plaskett Correspondence, *Science and Culture* **84** (2018) 285.
[18] *Collected Scientific Papers of Meghnad Saha*, CSIR, 1969.
[19] A.S. Eddington, Further Notes on the Radiative Equilibrium of the Stars, *Monthly Notic. Roy. Astron. Soc*. 77 (1917) 596.
[20] W. Nernst, *Die theoretischen und experimentellen Grundlagen des neuer Wärmesatzes*, Halle, 1918.
[21] M.N. Saha, Ionisation in the solar chromosphere, *Phil. Mag,* **40** (1920) 472.
[22] P.F. Chen, Coronal Mass Ejections: Models and Their Observational Basis, *Living Reviews in Solar Physics* **8** (2011) 1.
[23] M. Mukerjee, *Churchill's Secret War: The British Empire and the Ravaging of India during World War II*, Basic Books, 2011.
[24] M.N. Saha, Elements in the Sun, *Phil. Mag.* **40** (1920) 809.
[25] M.N. Saha, On the problems of temperature radiation of gases, *Phil. Mag.* **41** (1921) 267.
[26] M.N. Saha, On a physical theory of stellar spectra, *Proc. Roy. Soc*. **A99** (1921) 135.
[27] N. Lockyer, On the chemistry of the hottest stars, *Proc. Roy. Soc.* **61** (1897) 148.
[28] A.R. Choudhuri, *Nature's Third Cycle: A Story of Sunspots*, Oxford University Press, 2015.
[29] M.N. Saha, On electron-chemistry and its application to problems of radiation and astrophysics, *J. Asro. Soc. Ind.* **10** (1920) 72.
[30] H.N. Russell, The theory of ionization and the sun-spot spectrum, *Astrophys. J.* **55** (1922) 119.
[31] R.H. Fowler and E.A. Milne, The intensities of absorption lines in the stellar spectra, and the temperatures and pressures in the reversing layers of stars, *Monthly Notic. Roy. Astron. Soc*. **83** (1923) 403.
[32] S. Rosseland, *Theoretical Astrophysics*, Clarendon Press, 1936, p. xvi.
[33] R.H. Fowler, Dissociation-equilibria by the method of partitions, *Phil. Mag*. **45** (1923) 1.
[34] M.N. Saha, N.K. Sur and K. Mazumdar, Über einen experimentellen Nachweiss der thermischen Ionizierung der Elemente, *Zeitschrift für Physik* **40** (1927) 648.
[35] J. Evershed, Radial movement in sunspots, *Kodaikanal Observatory Bulletin* XV (1909).
[36] The Royal Society Archives.
[37] A.R. Choudhuri, FRS Election as a Recognition for Scientists of Colonial India, *Indian Journal of History of Science* (2018) in press.
[38] R. Singh and F. Riess, C.V. Raman, M.N. Saha and the Nobel Prize for the year 1930, *Indian Journal of History of Science* 34 (1999) 61.
[39] R.S. Anderson, *Nucleus and Nation*, University of Chicago Press, 2010.
[40] M.N. Saha, On physical theory of the solar corona, *Proc. Nat. Inst. Sci. Ind.* **8** (1942) 99.
[41] A.R. Choudhuri, *Astrophysics for Physicists*, Cambridge University Press, 2010, p. 32.
[42] A.R. Choudhuri, Practising Western Science outside the West: Personal Observations on the Indian Scene, *Social Studies of Science* **15** (1985) 475.